\documentclass[reprint,showpacs,amsmath,amssymb,aps,superscriptaddress,twocols]{revtex4-1}
\usepackage{amsmath}
\usepackage[tight,TABTOPCAP]{subfigure}
\usepackage[]{SIunits}
\usepackage[usenames, dvipsnames]{color}
\usepackage{natbib}
\usepackage[version=3]{mhchem} 
\usepackage[T1]{fontenc}       
\usepackage{hyperref}
\usepackage{soul}
\usepackage{notes2bib}
\usepackage{graphicx}

\begin{document}

\author{Maziar Heidari}
\affiliation{Max Planck Institute for Polymer Research, Ackermannweg 10, 55128 Mainz, Germany}
\author{Kurt Kremer}
\affiliation{Max Planck Institute for Polymer Research, Ackermannweg 10, 55128 Mainz, Germany}
\author{Robinson Cortes-Huerto}
\email{corteshu@mpip-mainz.mpg.de}
\affiliation{Max Planck Institute for Polymer Research, Ackermannweg 10, 55128 Mainz, Germany}
\author{Raffaello Potestio}
\email{raffaello.potestio@unitn.it}
\affiliation{Max Planck Institute for Polymer Research, Ackermannweg 10, 55128 Mainz, Germany}
\affiliation{Physics Department, University of Trento, via Sommarive, 14 I-38123 Trento, Italy}
\affiliation{INFN-TIFPA, Trento Institute for Fundamental Physics and Applications, I-38123 Trento, Italy}


\title{Spatially Resolved Thermodynamic Integration: An Efficient Method to Compute Chemical Potentials of Dense Fluids}

\date{\today}


\begin{abstract}
Many popular methods for the calculation of chemical potentials rely on the insertion of test particles into the target system. In the case of liquids and liquid mixtures, this procedure increases in difficulty upon increasing density or concentration, and the use of sophisticated enhanced sampling techniques becomes inevitable. In this work we propose an alternative strategy, spatially resolved thermodynamic integration, or SPARTIAN for short. Here, molecules are described with atomistic resolution in a simulation subregion, and as ideal gas particles in a larger reservoir. All molecules are free to diffuse between subdomains adapting their resolution on the fly. To enforce a uniform density profile across the simulation box, a single-molecule external potential is computed, applied, and identified with the difference in chemical potential between the two resolutions. Since the reservoir is represented as an ideal gas bath, this difference exactly amounts to the excess chemical potential of the target system. The present approach surpasses the high density/concentration limitation of particle insertion methods because the ideal gas molecules entering the target system region spontaneously adapt to the local environment. The ideal gas representation contributes negligibly to the computational cost of the simulation, thus allowing one to make use of large reservoirs at minimal expenses. The method has been validated by computing excess chemical potentials for pure Lennard-Jones liquids and mixtures, SPC and SPC/E liquid water, and aqueous solutions of sodium chloride. The reported results well reproduce literature data for these systems.
\end{abstract}

\maketitle
\makeatletter
\let\toc@pre\relax
\let\toc@post\relax
\let\toc@pre\relax
\makeatother

\setlength{\parindent}{0pt}

\section{Introduction}

An accurate estimation of the chemical potential ($\mu$) is essential to understand many physical and chemical phenomena \cite{Job-Herrmann-2006,Baierlein-2001}. Consider the study of nucleation processes at the nanoscale as an  example: in this context, prototypical systems such as water--alcohol mixtures \cite{Ceriotti2016}, mineral clusters \cite{Raiteri_2010,Demichelis_2011,DeLaPierre_2017} or ions in solution \cite{JACS137-13352-2015,Ferrario2002} present a challenge to existing computational methods. Even the computation of $\mu$ for aqueous table salt is still the subject of intense discussion \cite{JCP136_244508_2012,JCP144_124504_2016,JCP145_154111_2016,JCP143_044505_2015,JCP142_044507_2015,JCP139_124505_2013}.

Because of this, there has been a continuous, decades long effort to compute free energy differences and, in particular, chemical potentials \cite{DALY20122054,Kofke-Cummings-1997}. Given a molecular liquid, the free energy difference between a state of $N$ and one of $N+1$ molecules yields the chemical potential of the substance. There are several methods that implement this strategy, which can be classified \cite{Kofke-Cummings-1997} in expanded ensembles \cite{Nezbeda1991}, histogram-reweighting \cite{PhysRevLett.61.2635,PhysRevLett.63.1195,PhysRevE.51.5092} and, more important for the present discussion, free energy perturbation methods \cite{Widom,BENNETT1976245,PRL91-140601-2003,JCP122-144107-2005,JCTC7-4115-2011} and thermodynamic integration (TI) \cite{KTI}. 

\begin{figure}[h]
\includegraphics[width=\columnwidth,angle=0]{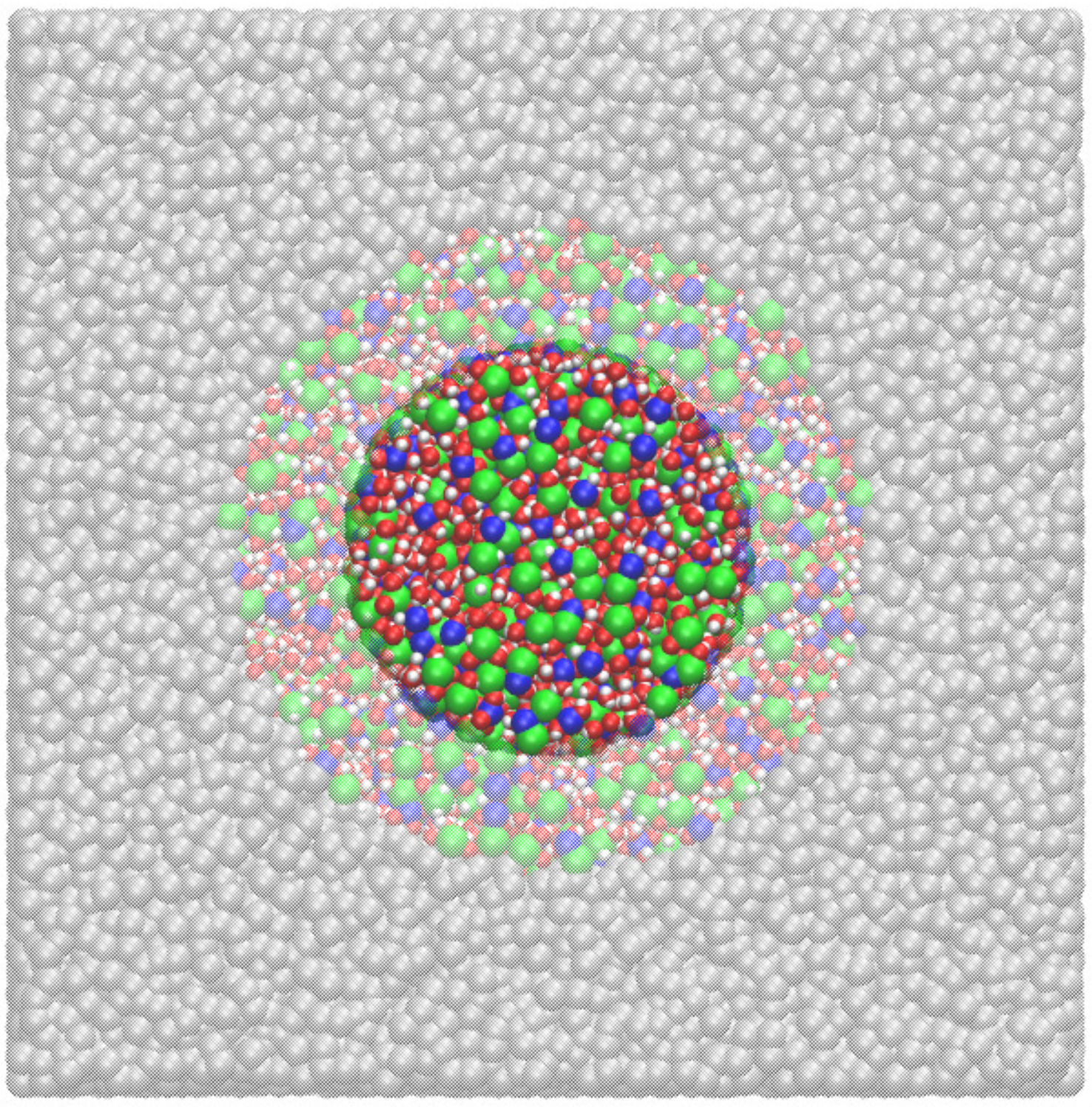}
 \caption{Simulation snapshot of a typical {\tt{H-AdResS}} setup of a system composed of sodium chloride in aqueous solution. Sodium, chlorine, oxygen and hydrogen atoms are represented by blue, green, red and white spheres, respectively. The atomistic, hybrid and ideal gas domains of the system are separated radially from the center of the simulation box, and the following convention is used to distinguish them in the figure: opaque (atomistic), transparent (hybrid) and silver (ideal gas) regions. The resolution of the molecules is determined by the switching function $\lambda (R)$ (Eq. \ref{eq:switching_func:01}).} \label{Model}
\end{figure}

Free energy perturbation methods are based on computing the Zwanzig identity \cite{Zwanzig}, which relates the target free energy difference to a canonical ensemble average over configurations generated by the $N$-particle Hamiltonian. A single stage application of Zwanzig identity results in the Widom method \cite{Widom} where frequent test particle insertions are used to calculate free energy differences. To increase the sampling efficiency, multi-stage applications of the Zwanzig identity, e.g Bennett acceptance ratio method (BAR)\cite{BENNETT1976245,PRL91-140601-2003,JCP122-144107-2005,JCTC7-4115-2011}, have been developed and are routinely used for simulations involving dense molecular fluids. Because of the necessity to sample a sufficiently large number of trial moves, these methods require a substantial computational effort that increases with density and/or concentration.
 Moreover, an adequate treatment of systems composed of complex molecules should include several intermediate states \cite{Kofke-Cummings-1997} or, in general, involve more sophisticated sampling techniques \cite{Perego2016}.

Thermodynamic integration \cite{KTI} is perhaps the most widely used method to compute free energy differences. TI is a dependable and powerful strategy which allows one to treat rather challenging systems such as solids \cite{Frenkel-Ladd-1984}, molecular crystals \cite{PhysRevLett.117.115702}, and molecular fluids \cite{Elio}. TI relies on the connection between the reference and the target states through a continuum of intermediates, parametrized by a factor combining the Hamiltonians of the two systems. The difference in free energy is obtained from ensemble averages of the appropriate observables computed on such intermediate states. Also in this case, the accuracy of the results depends on generating a sufficient number of system configurations -- most of them uninteresting -- thus hindering the overall efficiency of the method.

It is thus fair to say that the calculation of chemical potentials of dense liquids and complex molecular mixtures remains a challenging task. For these systems, the particle insertion procedure is highly inefficient and in some cases becomes unfeasible. To fill this gap, in this work we introduce a method to compute directly the excess chemical potential of a target system based on the Hamiltonian adaptive resolution scheme {\tt H-AdResS} \cite{Potestio2013Hamiltonian,Potestio2013Monte}. In this version of {\tt H-AdResS}, the target atomistic system (AT) is coupled to an ideal gas (IG) bath of point-like particles \cite{DellagoIG,kreis_EPJST2015}. The excess chemical potential is obtained by integrating in space the compensation forces necessary to ensure a uniform density profile throughout the whole AT$+$IG system; therefore we dub the method {\it spatially resolved thermodynamic integration}, or {\tt SPARTIAN}. {\tt SPARTIAN} is already implemented in a local version of the LAMMPS simulation package \cite{LAMMPS} and is made freely available\footnote{A in-house version of the LAMMPS package featuring all method implementations discussed in the present work can be freely downloaded from the webpage \url{http://www2.mpip-mainz.mpg.de/~potestio/software.php}.}.

{\tt SPARTIAN} is reminiscent of methods to compute the chemical potential in which inhomogeneities are imposed on the target system \cite{JCP101_7804_1994,JCP134_114514_2011,*JCP136_164503_2012}, or strategies in which the target and reference systems are physically separated by a semi-permeable membrane \cite{ROWLEY1995159}. In contrast with such methods, thermodynamic equilibrium is carefully monitored and identified with a uniform density profile across the simulation box. Moreover, finite size effects are made negligible when a substantially large reservoir is coupled to the atomistic region without increasing the computational cost -- as it is the case with IG particles since they do not interact. A similar idea was proposed in the context of force-based adaptive resolution simulations \cite{DelleSiteChemPot} where the calculation of effective potential energies is based on the configurations generated by a non-conservative dynamics. Conversely, our method relies on the same Hamiltonian function for both the generation of the dynamics and the computation of the chemical potential, the latter naturally emerging from the formulation of the {\tt H-AdResS} method, as discussed later on. Furthermore, {\tt SPARTIAN} is particularly efficient, because it employs IG particles in the reservoir, and it is flexible because its extension to multicomponent systems is straightforward.

The paper is organised as follows: in the {\it Method} section, after shortly describing {\tt H-AdResS}, the theoretical basis of {\tt SPARTIAN} is introduced. In the {\it Results and Discussion} section, excess chemical potential calculations are presented for Lennard-Jones liquids and liquid mixtures, as well as for SPC and SPC/E \cite{SPC1,*SPC2,*SPC3} water and for the Joung and Cheatham (JC) sodium chloride force field in SPC/E water \cite{JCFF}. The {\it Conclusion} section recapitulates the presented work.

\section*{Method}

\label{method}

\begin{figure}[h]
\centering
\includegraphics[scale=0.17,angle=0]{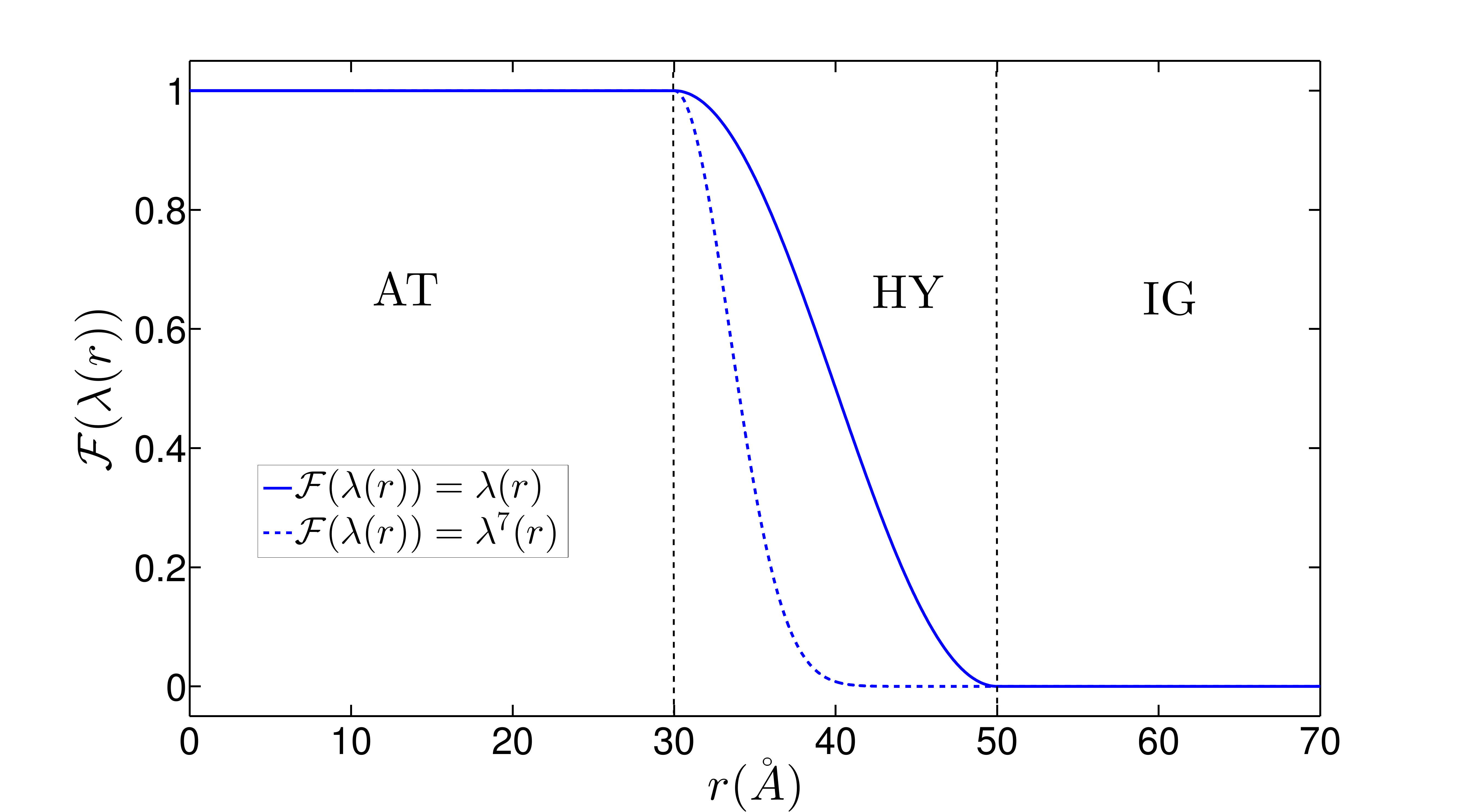}
\caption{Plot of the switching function, re-defined as $\mathcal{F}(\lambda (r))=\lambda^\nu (r) (\nu \ge 1)$ (see main text), as a function of the radial distance from the center of a cubic simulation box. Two different exponents, $\nu=1$ (see Eq. \ref{eq:switching_func:01}) and $7$, have been considered. In Eqs. \ref{hadress_H}-\ref{eq:deltaH_R} the function $\lambda$, hence with $\nu=1$, is directly employed to lighten the notation. In this plot, the sizes of the atomistic and hybrid regions are $r_{at}=30 \mbox{\AA}$, $d_{hy}=20 \mbox{\AA}$, respectively.} \label{lambda}
\end{figure}

One of the biggest challenges in computational soft matter physics is, arguably, to treat accurately and efficiently the wide range of time and length scales typically encountered when simulating complex molecular systems. One possibility to overcome this problem, in contrast to classical force-fields, consists of using coarse-grained models to access time and length scales usually out-of-range for fully atomistic simulations. However, there are situations in which it is convenient to keep a higher level of detail in a relatively small region within the simulation box (for example when the involved detailed chemistry is relevant) and simultaneously treat the neighbouring region using a computationally more efficient, i.e. coarse-grained, model. Adaptive resolution simulation methods \cite{adress1,adress2,adress3,annurev,adresstoluene,FritschPRL,Potestio2013Hamiltonian,Potestio2013Monte} implemented this strategy as schematically depicted in Fig. \ref{Model}. Specifically, in the {\tt H-AdResS} framework molecular interactions are treated either at the atomistic level in the AT region, 
fully coarse-grained in the CG region, or as an interpolation of the two in the HY region, in terms of a global Hamiltonian of the form:
\begin{eqnarray}\label{hadress_H}
&&H = \mathcal K + V^{int} + \sum_{\alpha} \left\{{\lambda_\alpha} {V^{AT}_\alpha} + {(1 - \lambda_\alpha)} {V^{CG}_\alpha} \right\}\,
\end{eqnarray}
with $\mathcal K$ the total kinetic energy and $V^{int}$ the sum of all intramolecular bonded interactions. The molecule $\alpha$ has resolution given by $\lambda_{\alpha}=\lambda(\mathbf{R}_{\alpha})$ associated to the center of mass coordinates $\mathbf{R}_{\alpha}$. The resolution of a molecule is thus determined by the value of this position-dependent switching function $\lambda_{\alpha}$, taking value 0 in the coarse-grained (CG) region and 1 in the fully atomistic (AT) region, which smoothly interpolates between such values in an intermediate hybrid (HY) region. As illustrated in Fig. \ref{Model}, the geometry of the AT region corresponds to a sphere of radius $r_{at}$, centered at the origin of coordinates. The HY region is a spherical shell of thickness $d_{hy}$ enclosing the AT region. The switching function plotted in Fig. \ref{lambda} depends on such a geometry, and it is defined by a function of the form:
\begin{eqnarray}\label{eq:switching_func:01}
\lambda(r)=\left\{
\begin{array}{ll}
1 & r \leq r_{at}\\
\cos^2\left(\frac{\pi(r-r_{at})}{2d_{hy}}\right) & {r_{at}}< r \leq r_{at}+d_{hy}\\
0 & r > r_{at}+d_{hy}
\end{array}
\right. .
\end{eqnarray}

Non-bonded molecular interactions are described at atomistic or coarse-grained level, and the resulting potential energy contribution for a given molecule $\alpha$ is the result of a weighted sum of two terms, $V^{AT}$ and $V^{CG}$, defined as:
\begin{eqnarray}\label{hadress_V}
&& V^{AT}_\alpha \equiv \displaystyle\frac{1}{2}\sum_{\beta,\beta\neq \alpha}^{N} \sum_{ij} V^{AT}(|\textbf{r}_{\alpha i} - \textbf{r}_{\beta j}|)\, ,\\ \nonumber
&& V^{CG}_\alpha \equiv \displaystyle\frac{1}{2}\sum_{\beta,\beta\neq \alpha}^{N} V^{CG}(|\textbf{R}_\alpha - \textbf{R}_\beta|)\, .
\end{eqnarray}

Note that there is no constraint regarding the use of arbitrary, e.g. many body, potentials. However, to lighten the notation, we carry out the following discussion making use of pairwise interactions. The total force acting on atom $i$ of molecule $\alpha$ is given by:
\begin{eqnarray}\label{hadress_F}
\textbf{F}_{\alpha i} &=& \textbf{F}^{int}_{\alpha i}\\ \nonumber
&+& \sum_{\beta,\beta\neq \alpha} \left\{ \frac{\lambda_\alpha + \lambda_\beta}{2} \textbf{F}^{AT}_{\alpha i|\beta} + \left(1 - \frac{\lambda_\alpha + \lambda_\beta}{2}\right) \textbf{F}^{CG}_{\alpha i |\beta} \right\}\\ \nonumber
&-& \left[ V^{AT}_\alpha - V^{CG}_\alpha \right] \nabla_{\alpha i}\lambda_\alpha\,
\end{eqnarray}
with $\textbf{F}^{int}_{\alpha i}$ the intramolecular total force on atom $i$ of molecule $\alpha$. The second term is the sum over all molecules $\beta\neq \alpha$ within cutoff distance of AT and CG forces weighted by the average resolution of the molecule pair ($\alpha$, $\beta$). The origin of the last term in the sum can be traced to the fact that molecules interact depending on their position within the simulation box. This breaking of translational invariance generates a force that acts on molecules located in the HY region, where $\nabla \lambda \ne 0$, and pushes them towards the AT or CG regions depending on the sign of $(V^{AT}_\alpha - V^{CG}_\alpha)$.
 
This drift force thus contributes to a pressure imbalance in the HY region. Furthermore, by joining AT and CG representations of a system using open boundaries, particles will diffuse to stabilise differences in pressure and chemical potential between the two representations. Hence, a non homogeneous density profile appears as a result of molecules being pushed to the region with lower molar Gibbs free energy \cite{Potestio2013Hamiltonian,Potestio2013Monte}. To impose a uniform density profile, an extra term is included in the Hamiltonian:
\begin{equation}
H_{\Delta} = H -\sum_{\alpha=1}^{N} \Delta H(\lambda(\mathbf{R}_{\alpha}))\, ,
\end{equation}
that compensates on average the drift force discussed above and imposes the pressure at which the AT and IG models exhibit the same density. To compensate the drift force, $\Delta H(\lambda(\mathbf{R}_{\alpha}))$ should satisfy the relation:
\begin{eqnarray}
&&\frac{d \Delta H(\lambda)}{d \lambda}\biggl|_{\lambda = \lambda_\alpha} \equiv \mathcal V(\lambda_\alpha) = 
\left\langle  \left[ V^{AT}_\alpha - V^{CG}_\alpha \right] \right\rangle_{{\bf R}_\alpha}\, ,
\end{eqnarray}
in such a way that the total drift force becomes:
\begin{eqnarray} 
\hat{\textbf{F}}^{dr}_\alpha = \left(V^{AT}_\alpha - V^{CG}_\alpha - 
\mathcal V(\lambda_\alpha)\right) 
\nabla \lambda \left(\textbf{R}_\alpha\right)\,
\end{eqnarray}
where $\langle \hat{\textbf{F}}^{dr}_\alpha\rangle \equiv 0$. The strategy introduced in Ref. \citenum{Pep2015Stat,Heidari2016} is used to compute $\mathcal V(\lambda_\alpha)$. This method, whose most technical aspects are detailed in the SI, averages over short time intervals the drift force acting upon each molecule species in the hybrid region as a function of the resolution. In between intervals, the computed average is used to update the drift force compensation $\mathcal V(\lambda)$, in such a way that correlations between molecules with different instantaneous resolutions are explicitly taken into account.

As anticipated, the drift force is not the only source of density imbalance in the system. The models coupled together in the same setup naturally feature different pressures, and a non-uniform density profile emerges as a consequence of the tendency of the system to equate the pressure imbalance between the two subdomains. To compensate for this density gradient it is customary to introduce a force, dubbed thermodynamic force \cite{FritschPRL,Heidari2016}, which, just as the aforementioned free energy compensation, acts only on the molecules in the HY region. This force is obtained through an iterative procedure, with updates proportional to the density gradient:
\begin{eqnarray}\label{thermo_force}
\textbf{F}_{n+1}^{th} = \textbf{F}_{n}^{th} + \frac{c \nabla \rho_n(x)}{\rho^*}\, .
\end{eqnarray}

\begin{figure}[h]
\centering
\includegraphics[scale=0.2,angle=0]{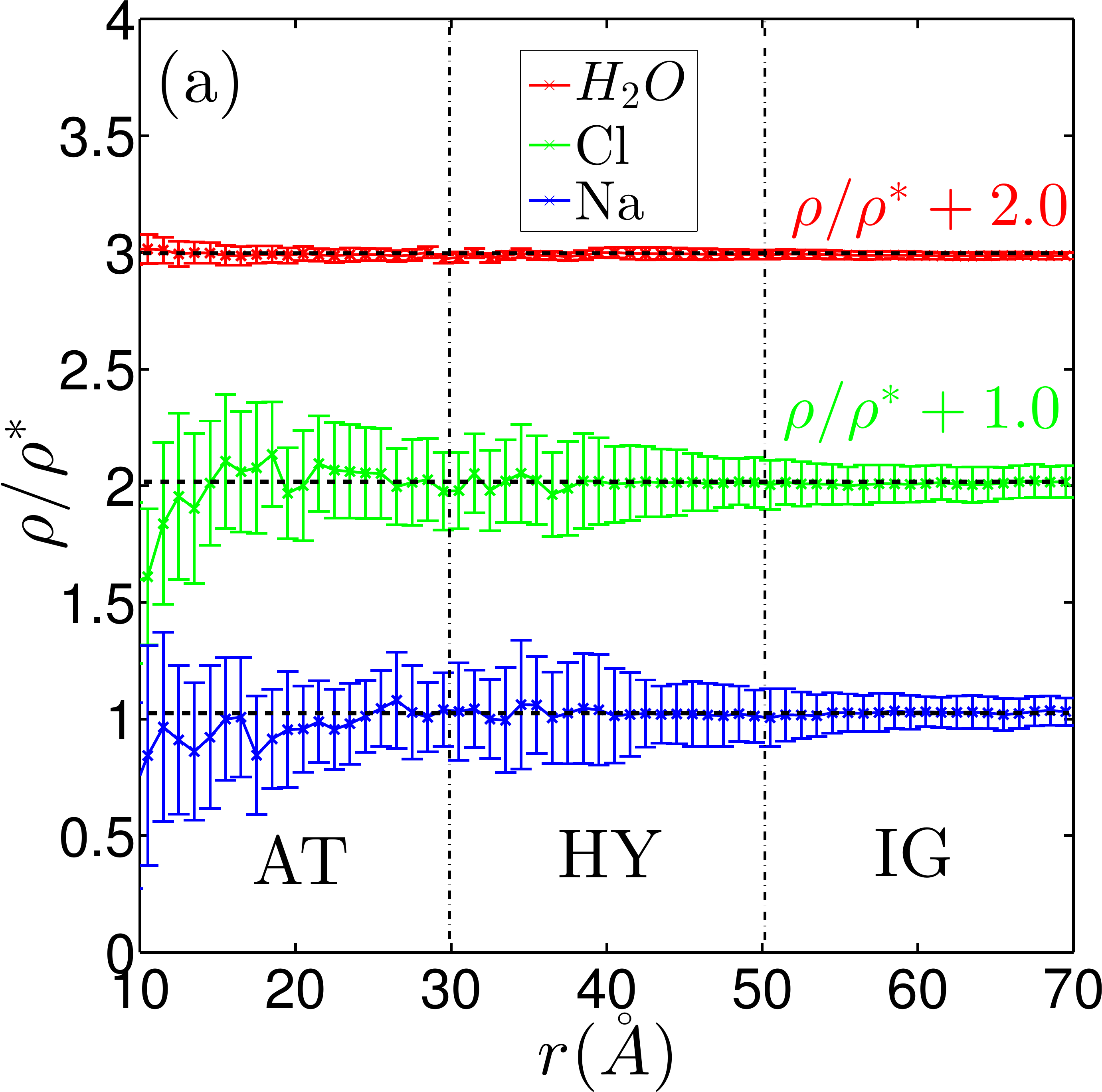}\\
\includegraphics[scale=0.2,angle=0]{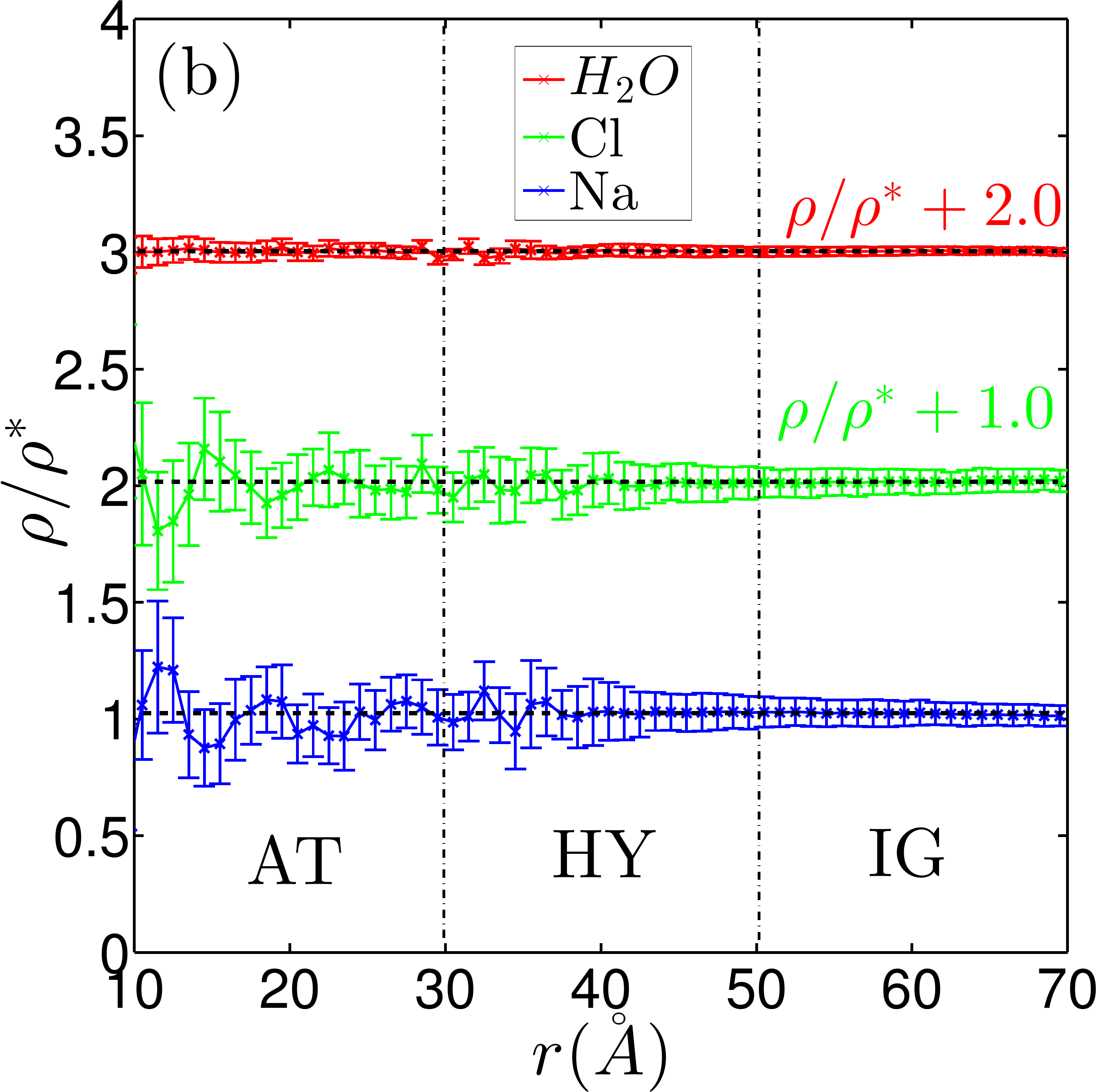}
\caption{Normalized and shifted density profiles of sodium chloride solutions as a function of the radial distance from the 
origin of the spherical atomistic region, for molalities $m=3.0$ (a) and $m=7.0$ (b). The increasing error bar sizes for distances approaching the origin are consistent with the reduced number of molecules available in small spherical shells.} \label{dens}
\end{figure}

The parameter $c$ modulates the force strength and has units of energy, $\rho^{*}$ is the reference density, and $\rho_{n}$ is the density profile computed at the $n$-th step of the iteration. This procedure converges to a uniform density profile throughout the simulation box when $\nabla \rho = 0$. In Fig. \ref{dens}, converged density profiles for {\tt H-AdResS} simulations of sodium chloride with the JC force field in SPC/E water \cite{JCFF} are presented to illustrate this point.
 
The total force acting on the molecules in the hybrid region is the sum of the compensation needed to cancel the drift force plus the thermodynamic force, hence:
\begin{equation}
\frac{d \Delta H(\lambda)}{d \lambda}\biggl|_{\lambda = \lambda_\alpha} = - \mathcal V(\lambda_\alpha) \nabla \lambda (\textbf{R}_\alpha) + \textbf{F}^{th}_\alpha
\end{equation}
from which we obtain:
\begin{equation} \label{eq:deltaH_R}
\Delta H(R_b) = - \int_{R_a}^{R_b} dR \left [ - \mathcal V(\lambda) \nabla \lambda (R) + \textbf{F}^{th}(R) \right]\, ,
\end{equation}
where $R_{a}=r_{at}+d_{hy}$ and $R_{b}=r_{at}$. Eq. \ref{eq:deltaH_R} allows one to compute the free energy compensation necessary to the system to attain a uniform density profile. 

The compensation term to the Hamiltonian has a simple and fundamental physical meaning, that is, it is the difference in chemical potential between the AT and CG regions \cite{Potestio2013Hamiltonian,Potestio2013Monte}:
\begin{equation}\label{eq:deltaH_deltaG_mu}
\Delta H(\lambda(R)) \equiv \frac{\Delta G(R)}{N} = \Delta \mu(R)\, ,
\end{equation}
with $\Delta G/N$ being the molar Gibbs free energy. Note that all quantities appearing in Eq. \ref{eq:deltaH_deltaG_mu} are functions of the molecule's position $R$: indeed, all free energies and chemical potential differences are computed with respect to a reference given by $R = R_a \equiv \lambda = 0$, that is, with respect to a CG model of arbitrary nature and complexity. If these functions are computed for $\lambda = 1$, one obtains the free energy / chemical potential difference between AT and CG model.

This is precisely the core of the method proposed here: in a nutshell, {\tt H-AdResS} is equivalent to a spatially resolved Kirkwood thermodynamic integration \cite{Potestio2013Hamiltonian,Potestio2013Monte}. Therefore, it is possible to calculate $\Delta \mu$ between a target (AT) and a reference (CG) system coexisting at the state ($\rho^{*}$,T) by varying $\lambda$, a coupling parameter of the global Hamiltonian of the system, across the interface.  
However, to compute the chemical potential of the target AT system, it is necessary to know the one of the reference CG system. To circumvent this extra step, we couple the AT target system to a bath of ideal gas (IG) particles \cite{kreis_EPJST2015}. In this case, the global Hamiltonian of the system reduces to:
\begin{eqnarray}\label{eq:hadress_H}
&&H = \mathcal K + V^{int} + \sum_{\alpha} {\mathcal F(\lambda_\alpha)} {V^{AT}_\alpha},
\end{eqnarray}
since $V^{CG}_{\alpha}\equiv 0$ $\forall \alpha$. We have introduced, in Eq. \ref{eq:hadress_H}, a modification to the switching function $\lambda$, which has been replaced by a function $\mathcal{F}(\lambda)=\lambda^{\nu}$ ($\nu \ge 1$). Similarly to $\lambda$, $\mathcal{F}(\lambda)$ takes values between 0 and 1, however it has a faster decay to zero as it approaches the IG region. This is required because it might happen that two molecules come extremely close  to each other when they are both located near the IG/HY interface. The choice $\nu=7$ is sufficient to smooth out such divergent interactions and avoid the systematic sampling of huge potential energy values which might affect the calculation of free energy compensations. Furthermore, as it is depicted in Fig. \ref{lambda}, by increasing the exponent $\nu$ the effective boundary between hybrid and CG region moves deeper towards the AT domain, leading to a more stable and controlled thermodynamic force convergence (see Eq. \ref{thermo_force}). In addition to this, the potential of the high resolution region is capped at a distance $\hat{r}$ to avoid large forces resulting from overlapping molecules:
\begin{eqnarray}\label{eq:fec:01}
V^{AT}(r)=\left\{
\begin{array}{ll}
V^{AT}(\hat{r})-\left.\frac{\partial V^{AT}}{\partial r}\right|_{r=\hat{r}}(r-\hat{r}) & r < \hat{r}\\
V^{AT}(r) & r \geq \hat{r}\\
\end{array}
\right. .
\end{eqnarray}

For the systems and simulation conditions considered here, these overlapping events are anyways rare (approximately one in 0.5 nanoseconds). However, for the sake of stability, they need to be identified and removed from the simulation. We verified that this capping does not change appreciably the thermodynamical nor the structural properties of the system. In particular, we performed fully atomistic simulations of SPC/E water with and without capping potentials and we do not observe any significant difference in the RDFs (data not shown). More important for the present discussion, capping the potential does not affect the calculated values of free energies and chemical potentials. The high energy contributions resulting from the excluded volume are located in the tail of the energy distribution of the system and have an accordingly small effect.

From the Hamiltonian given by \eqref{eq:hadress_H}, the total force acting on the atom $i$ of the molecule $\alpha$ whose resolution is $\lambda_{\alpha}$ is given by:
\begin{eqnarray}\label{hadress_F}
\textbf{F}_{\alpha i} &=& \textbf{F}^{int}_{\alpha i}\\ \nonumber
&+& \sum_{\beta,\beta\neq \alpha} \left\{ \frac{\mathcal F(\lambda_\alpha) + \mathcal F(\lambda_\beta)}{2} \textbf{F}^{AT}_{\alpha i|\beta} \right\}\\ \nonumber
&-& V^{AT}_\alpha  \frac{\partial \mathcal F}{\partial \lambda}|_{\lambda=\lambda_\alpha}\nabla_{\alpha i}\lambda_\alpha\, .
\end{eqnarray}

In addition to the obvious computational advantage of using an IG over a standard, i.e. interacting, CG model, by coupling an AT model with an IG bath of particles at a thermodynamic state with density $\rho^*$ and temperature $T$ it is possible to compute automatically the excess chemical potential from $\Delta H(\lambda(\mathbf{R}_{\alpha}))$. Moreover, in {\tt H-AdResS} this result can be immediately extended to multicomponent systems \cite{Potestio2013Monte}, thus:
\begin{equation}
\mu^{i}_{ex} = \Delta H_{i}(\lambda(\mathbf{R}_{i,\alpha}))\, , 
\end{equation}
where the index $i$ indicates the species in the mixture. This implies that in the case of a liquid mixture the compensations are computed for every species separately {\it yet simultaneously}, therefore in a single {\tt H-AdResS} equilibration it is possible to compute all the $\mu^{i}_{ex}$. 

We conclude this section by pointing out that the strategy presented here to compute the chemical potential of dense liquids stems entirely from an implicit property of adaptive resolution approaches, and in particular of the {\tt H-AdResS} method. As a matter of fact, the procedures to calculate the free energy compensations are a basic step and fundamental ingredient of this method, and are necessary in order to prepare the simulation setup with a uniform density profile.

\section*{Results and Discussion}
\label{results}

\subsection*{Lennard-Jones fluid and comparison with the Widom method}

We first validate our method by computing the excess chemical potential $\mu_{ex}$ of a Lennard-Jones liquid. We consider systems whose interaction potential is given by a (12,6) Lennard-Jones (LJ) potential truncated and shifted with cutoff radius 2.5$\sigma$. The chosen parameters are $\sigma=\epsilon=1$. The results for this section are expressed in LJ units with mass $m=1$, time $\tau=\sqrt{m\sigma^2/\epsilon}$, temperature $k_{B}T = 2\epsilon $ and pressure $\sigma^{3}/\epsilon$. Simulations were carried out using LAMMPS \cite{LAMMPS} with a time step $5\times 10^{-4}\tau$. Constant temperature was enforced by a Langevin thermostat with coupling parameter $100\tau$. A system of size $N=1687500$ was considered in the density range $\rho = 0.3\cdots 1.0$, with a corresponding number of particles in the AT region ranging in the $2800\cdots 9400$ interval. The radius of the atomistic region and the thickness of the hybrid shell are both 15 $\sigma$. We performed equilibration runs of $10^5$ MD steps and production runs of $10^6$ MD steps. Furthermore, to compare the results obtained with the method outlined here, we use the Widom method \cite{Widom} for equivalent systems but of  size N = 1000 and in the range of densities $\rho = 0.3\cdots 0.8$ 

\begin{figure}[h]
\includegraphics[scale=0.17,angle=0]{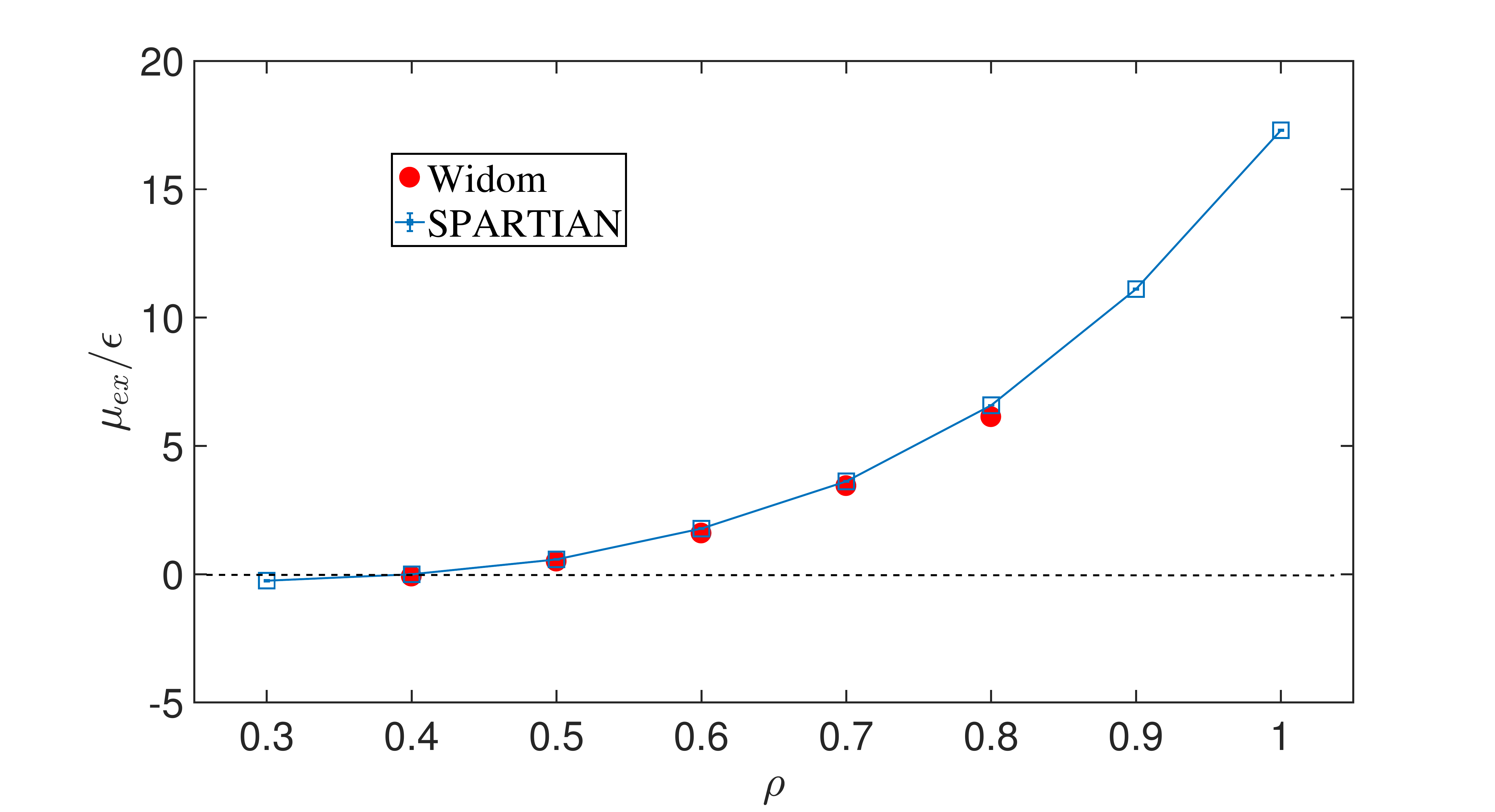}
\caption{Excess chemical potential for the pure Lennard-Jones liquid computed with the {\tt SPARTIAN} method (red full circles) and with the Widom method (blue open squares). In both cases, the error bars are smaller than the symbol size. Note that the {\tt SPARTIAN} method provides results consistent with the observed trend also in regions of density beyond the range of applicability of Widom method. The horizontal dashed line indicates the ideal gas case in which $\mu_{ex}=0$.} \label{Fig1}
\end{figure}

Results for $\mu_{ex}$ as a function of $\rho$ are shown in Fig. \ref{Fig1} where a remarkable agreement between the Widom and the {\tt SPARTIAN} method results can be appreciated. Furthermore, and in contrast to the Widom method, the adaptive resolution calculation of $\mu_{ex}$ provides consistent results for high densities. This result is expected since {\tt SPARTIAN} takes advantage of the accurate sampling made possible by the large density of the system. 
 
\subsection*{Lennard-Jones mixture}

To test the range of applicability of {\tt SPARTIAN}, we compute the excess chemical potential of simple molecular liquid mixtures. In particular, we simulate a glass-forming binary Lennard-Jones mixture using the interaction parameters of Ref. \citenum{KobAndersen}. The mixture consists of $80\%$ particle A and $20\%$ particle B. In terms of length and energy units of $\sigma_{AA}=1.0$ and $\epsilon_{AA}=1.0$, the interaction parameters are $\sigma_{BB}=0.88$, and $\sigma_{AB}=0.8$, $\epsilon_{BB}=0.5$, $\epsilon_{AB}=1.5$, and the cut-off radius and the temperature are set $r_c=2.5\sigma$ and $k_BT=0.75$, respectively. In this case as well, 
the radius of the atomistic region and the thickness of the hybrid shell is 15 $\sigma$.

As discussed in the Method sections, we can treat independently all the species in the mixture, i.e., there is a density profile associated to species A and B and thermodynamic forces are applied to every density profile. In this way, we can automatically extract the excess chemical potential for every species in the mixture.

\begin{table}[h]
\begin{center}\caption{Excess chemical potentials of both particle species of the Lennard-Jones mixture as of Ref. \citenum{KobAndersen} at temperature $T=0.75/k_B$, computed by the {\tt SPARTIAN} method and compared to the values obtained {\it via} particle insertion enhanced by means of Metadynamics \cite{Perego2016}. The unit of all values is $\epsilon_{AA}$.}
    \begin{tabular}{ c l c l c p{5cm} |} 
    \hline
    Component / model & {\tt SPARTIAN} & Ref. \citenum{Perego2016}   \\ \hline
    $\mu_{ex}^A$ &  3.95 $\pm$ 0.02 &3.99 $\pm$ 0.04\\ \hline
    $\mu_{ex}^B$ & -4.61 $\pm$ 0.06 & -4.65 $\pm$ 0.02 \\ \hline 
    \end{tabular}
\end{center}
\end{table}

Results for this system are presented in Table I, where an excellent agreement with calculations based on Metadynamics \cite{Perego2016} is apparent. The sign of the excess chemical potential reflects that it is more favorable to insert the small B particles in the system due to their low concentration and relatively (with respect to A particles) weak interaction energy. An interesting behaviour can be observed in the errors: our method, in fact, provides more accurate estimates of the excess chemical potential of A particles rather than B particles. This behavior differs from that of methods using test particle insertions, and reflects the fact that {\tt SPARTIAN} substantially relies on -- and takes advantage of -- accumulating statistics, which improves as the mole fraction of solute molecules increases.

\subsection*{SPC/E Water}

The calculation of the chemical potential for dense liquid water is a rather challenging task. Standard methods to compute free energy differences like the Widom insertion do not converge convincingly \cite{DALY20122054} and it is thus necessary to use more sophisticated methods even for this system composed of relatively small molecules. 

We have computed the excess chemical potential of two rather popular water models, SPC and SPC/E \cite{SPC1,*SPC2,*SPC3}. Molecular dynamics simulations have been carried out for 117000 water molecules. The size of the cubic simulation box is $152 \mbox{\AA}$, the radius of the AT region is $30 \mbox{\AA}$ and the thickness of the HY region is $20 \mbox{\AA}$, which is larger than the Bjerrum length of pure water ($\lambda_B =7.5\AA)$. A 0.5 ns equilibration run has been performed with time step $\delta t = 1$ fs in the NPT ensemble for a fully atomistic system. Temperature and pressure are enforced at $T=$298 K and $P=$1 bar using the Nos\'e-Hoover thermostat and barostat with damping coefficient of 100 fs and 1000 fs, respectively. This procedure provides the initial configuration for the subsequent {\tt SPARTIAN} simulations, which have been performed in the NVT ensemble. Here we have used the same $\delta t$ and enforced the same temperature $T$ with a Langevin thermostat with coupling parameter $100\tau$.
 
Results are presented in Table II. For completeness, we have compared with results obtained using thermodynamic integration (TI) (SPC\cite{quintana}, SPC/E\cite{DelleSiteChemPot}) and two-stage particle insertion methods, i.e. Bennett acceptance ratio method (BAR) \cite{BENNETT1976245,PRL91-140601-2003,JCP122-144107-2005,JCTC7-4115-2011} (SPC\cite{Sauter_2016}, SPC/E\cite{JCP142_044507_2015}). Once again, our results agree reasonably well, approximately less than $5\%$ difference, with the values reported in the literature in all cases. 

\begin{table}[h]
\begin{center}\caption{Excess chemical potential of water molecules at temperature T=298 K as computed in this work, compared to the values obtained with thermodynamic integration \cite{quintana,DelleSiteChemPot} and two-stage particle insertion methods \cite{JCP142_044507_2015}. The experimental value is -26.46 kJ/mol \cite{BenNaim1984}. The unit of all values is kJ/mol.}
    \begin{tabular}{ c l c l c l c l c l c p{5cm} |} 
    \hline
    Water model & {\tt SPARTIAN} & TI & BAR  \\ \hline
    SPC &   -25.68 $\pm$ 0.02 & -23.9 $\pm$  0.6 \cite{quintana} & -26.13 $\pm$ 0.05 \cite{Sauter_2016} \\ \hline
    SPC/E &  -29.01 $\pm$ 0.09 & -29.53 $\pm$ 0.03 \cite{JCP142_044507_2015}& -29.70 $\pm$ 0.05 \cite{Sauter_2016}  \\ \hline 
    \end{tabular}
\end{center}
\end{table}

\subsection*{Aqueous solution of sodium chloride}

Lastly, we have computed the excess chemical potentials $\mu_{ex}^{NaCl}$ and $\mu_{ex}^{H_{2}O}$ for sodium chloride (NaCl) in water. For this prototypical electrolyte solution present in biological, geological, and industrial contexts, many computational studies have been devoted to calculate $\mu_{ex}$ using various different methodologies \cite{JCP136_244508_2012,JCP144_124504_2016,JCP145_154111_2016,JCP143_044505_2015,JCP142_044507_2015,JCP139_124505_2013}. This wealth of results constitutes an excellent database to benchmark the method proposed here. Furthermore, strong electrostatic interactions present in salt solutions provide us with a challenging testing ground.

We have performed molecular dynamics simulations for NaCl aqueous solutions with 117000 water molecules and in the range of molalities mol$_{\text{solute}}$/kg$_{\text{solvent}}= 0 \cdots 10$. This interval includes substantially higher ion concentrations than the ones reported recently \cite{JCP142_044507_2015,JCP144_124504_2016}. We have used the force field parameters of the $Na^{+}$ and $Cl^{-}$ ions from Ref. \citenum{JCFF}, truncated and shifted at $r_{\text{cutoff}}^{LJ}=10\AA$ (for the non--Coulombic terms), and the SPC/E \cite{SPC1,*SPC2,*SPC3} parameters for water. This combination provides the value of solubility closest to experimental measurements \cite{JCP136_244508_2012}. The cubic simulation box side is $154.5 \mbox{\AA}$, the radius of the AT region is $30 \mbox{\AA}$, and the thickness of the HY region is $20 \mbox{\AA}$. As previously done for pure water, we first performed a 1 ns long equilibration run for the fully atomistic system with $\delta t = 1$ fs in the NPT ensemble. We kept temperature and pressure constant at $T=$298 K and $P=$1 bar using the Nos\'e-Hoover thermostat and barostat with damping coefficient of 100 fs and 1000 fs, respectively. The resulting equilibrated configurations have been employed as starting point for {\tt SPARTIAN} simulations which have been performed in the NVT ensemble using the same $\delta t$ and $T$. We have controlled the temperature with a Langevin thermostat with coupling parameter $100\tau$.

\begin{figure}[h]
\centering
\includegraphics[scale=0.17,angle=0]{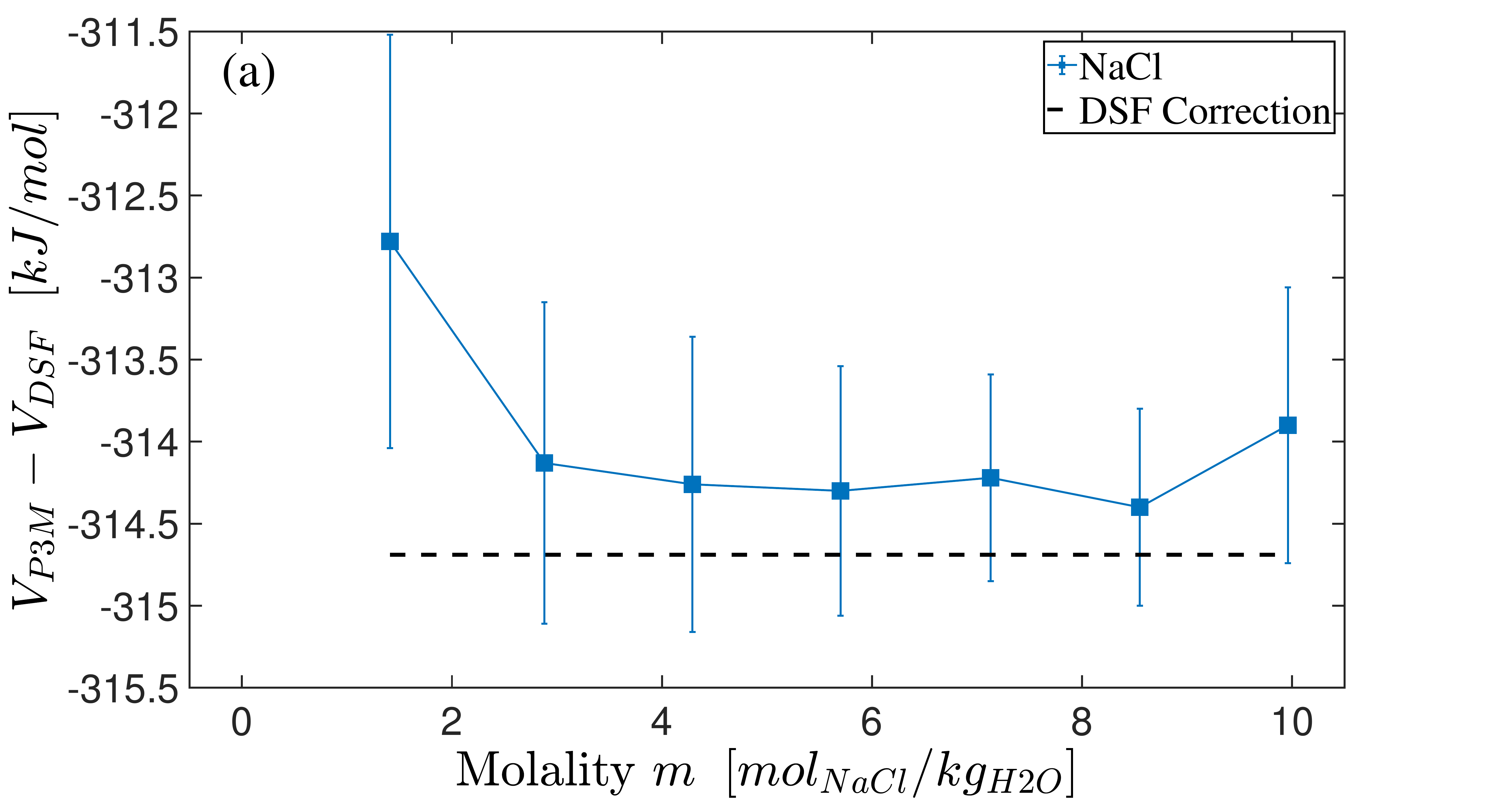}\\
\includegraphics[scale=0.17,angle=0]{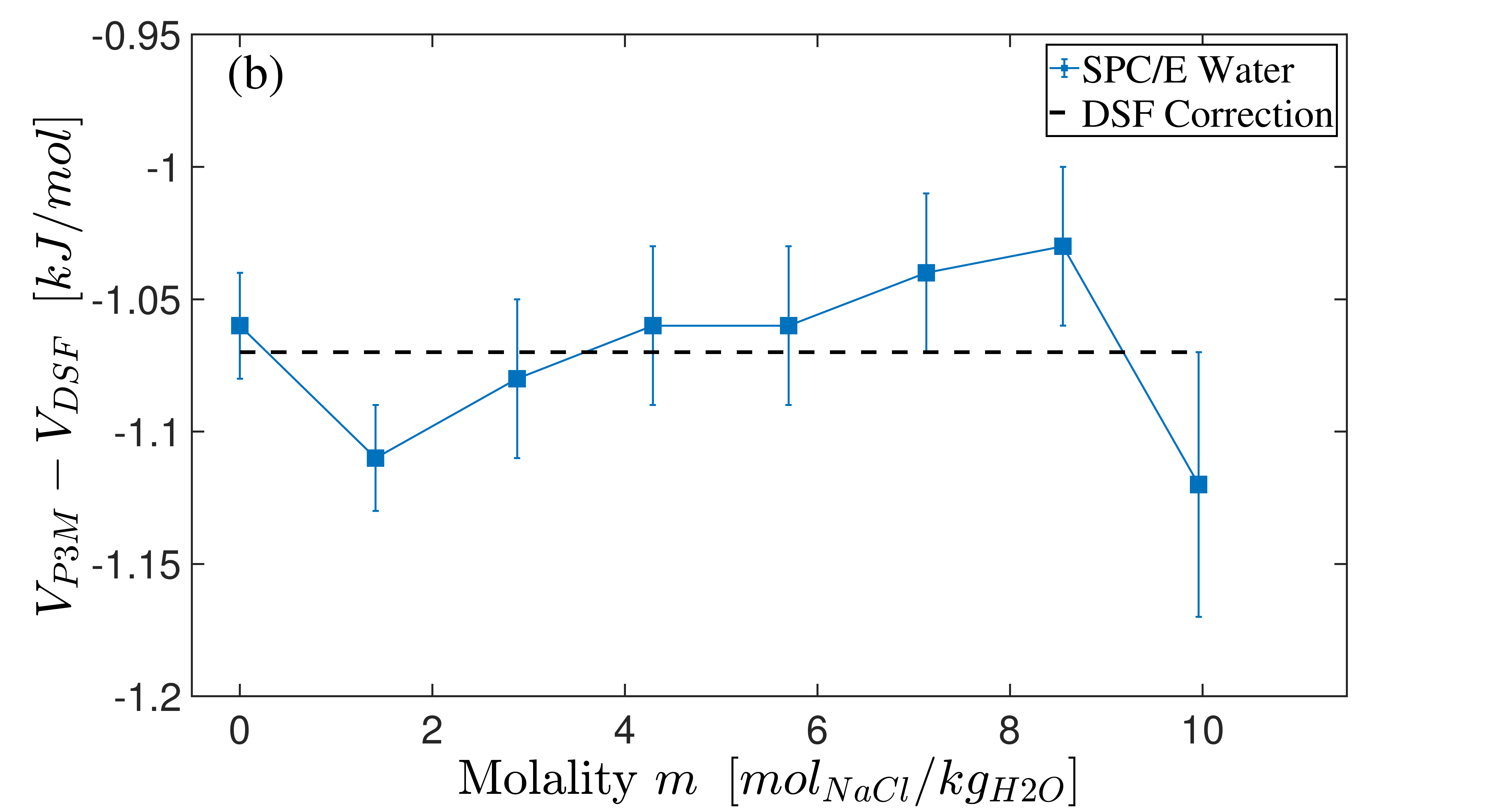}
\caption{Difference between the total electrostatic potential of an aqueous solution of NaCl as computed {\it via} Ewald summation method and DSF method, as a function of molality. For both NaCl (a) and water (b), the difference is nearly independent of the salt concentration, and its dominant contributions can be rationalized in terms of the sum of self-interaction ($V^{self}$) and excluded electrostatic interactions ($V^{excl}$) of the DSF potential, whose theoretical value is reported as dashed lines. For a detailed discussion of these corrections see the SI, in particular Eqs. (SI3) and (SI4).}\label{Fig2}
\end{figure}

The {\tt H-AdResS} method relies on the use of short-range potentials and forces to treat electrostatic interactions. In a previous study, we have implemented and validated the damped shifted force potential \cite{fennel_JCT2006} (DSF) in Hamiltonian adaptive resolution simulations \cite{Heidari2016}. Following Ref. \citenum{Heidari2016}, the DSF parameters employed in the present study are $\alpha = 0.2\AA^{-1}$ and $r_{\text{cutoff}}^{DSF}=12\AA$. Since we expect electrostatics to influence the accuracy of the chemical potential calculations, it is necessary to assess how different the results might be when using either the DSF method or the standard Ewald summation method \cite{Takahashi2013}. We have performed fully atomistic NPT simulations for the same setups previously discussed. We then compared the difference in electrostatic potential $V_{P3M}-V_{DSF}$ between the Ewald and DSF calculations for NaCl and water molecules. The results are presented in Fig. \ref{Fig2}, where a nearly constant difference is observed for all salt concentrations considered (the fluctuations about the average value of $V_{P3M}-V_{DSF}$ across all molalities are of approximately 1$\%$  and 3$\%$ for NaCl and water, respectively).
Since the difference in potential energies when using Ewald summation or DSF can be treated as constant, then $V_{P3M}-V_{DSF}$ for every species in the system amounts to a constant shift in the excess chemical potential. Furthermore, it is possible to avoid doing an extra simulation using the Ewald method if we investigate the theoretical origin of the mismatch with respect to simulations using the DSF method. The DSF potential includes a contribution $V^{self}$ that guarantees charge neutrality at a given cutoff radius, and a contribution $V^{excl}$ that excludes, in the case of rigid molecules such as SPC/E water, intramolecular electrostatic interactions (see SI for a detailed discussion). The sum of the two contributions is force field-- but not salt concentration--dependent, as evidenced by the black horizontal lines in Fig. \ref{Fig2}, and it accounts precisely, within statistical error, for the chemical potential shift $V_{P3M}-V_{DSF}$.

\begin{figure}[h]
\centering
\includegraphics[scale=0.17,angle=0]{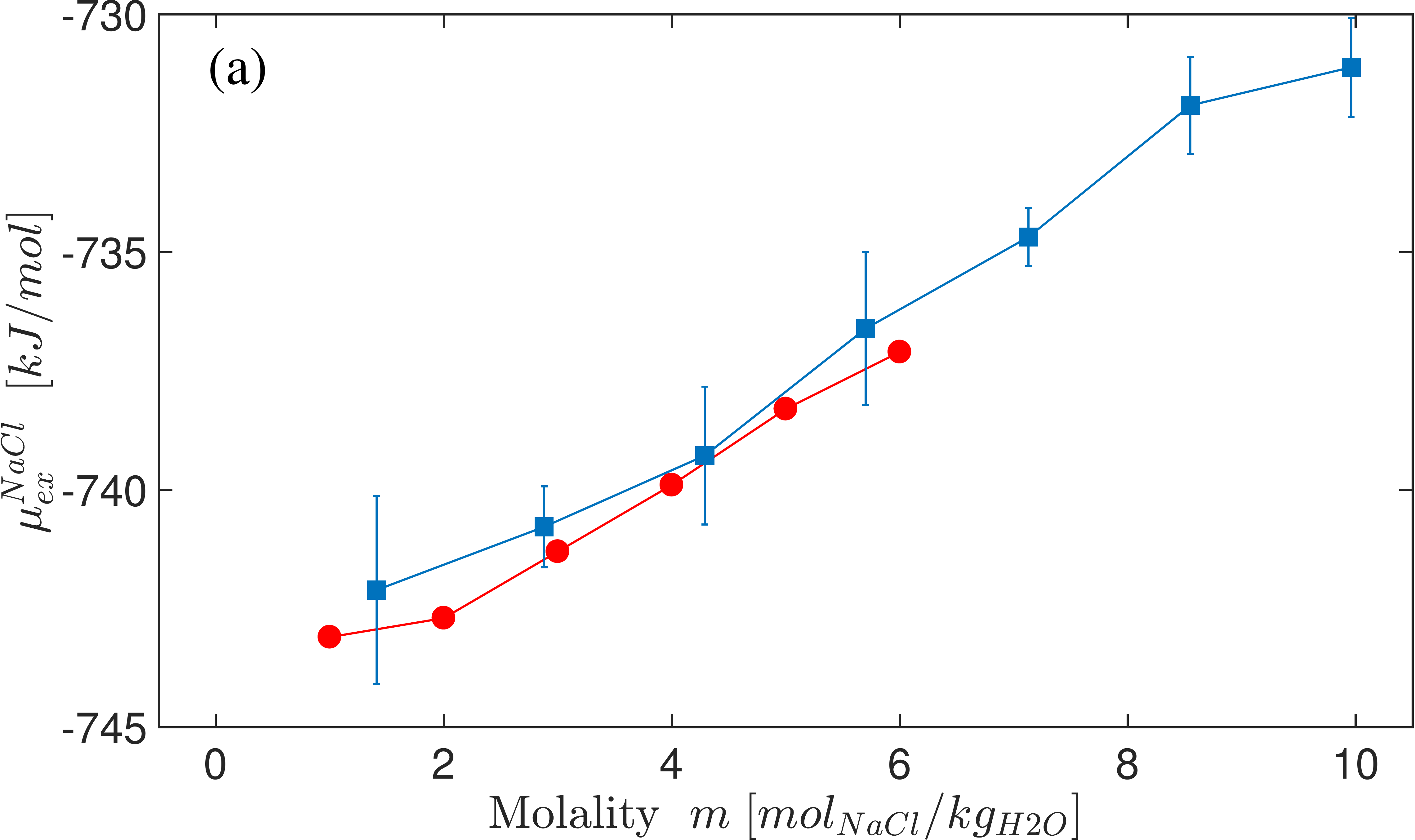}\\ \vspace{0.5cm}
\includegraphics[scale=0.17,angle=0]{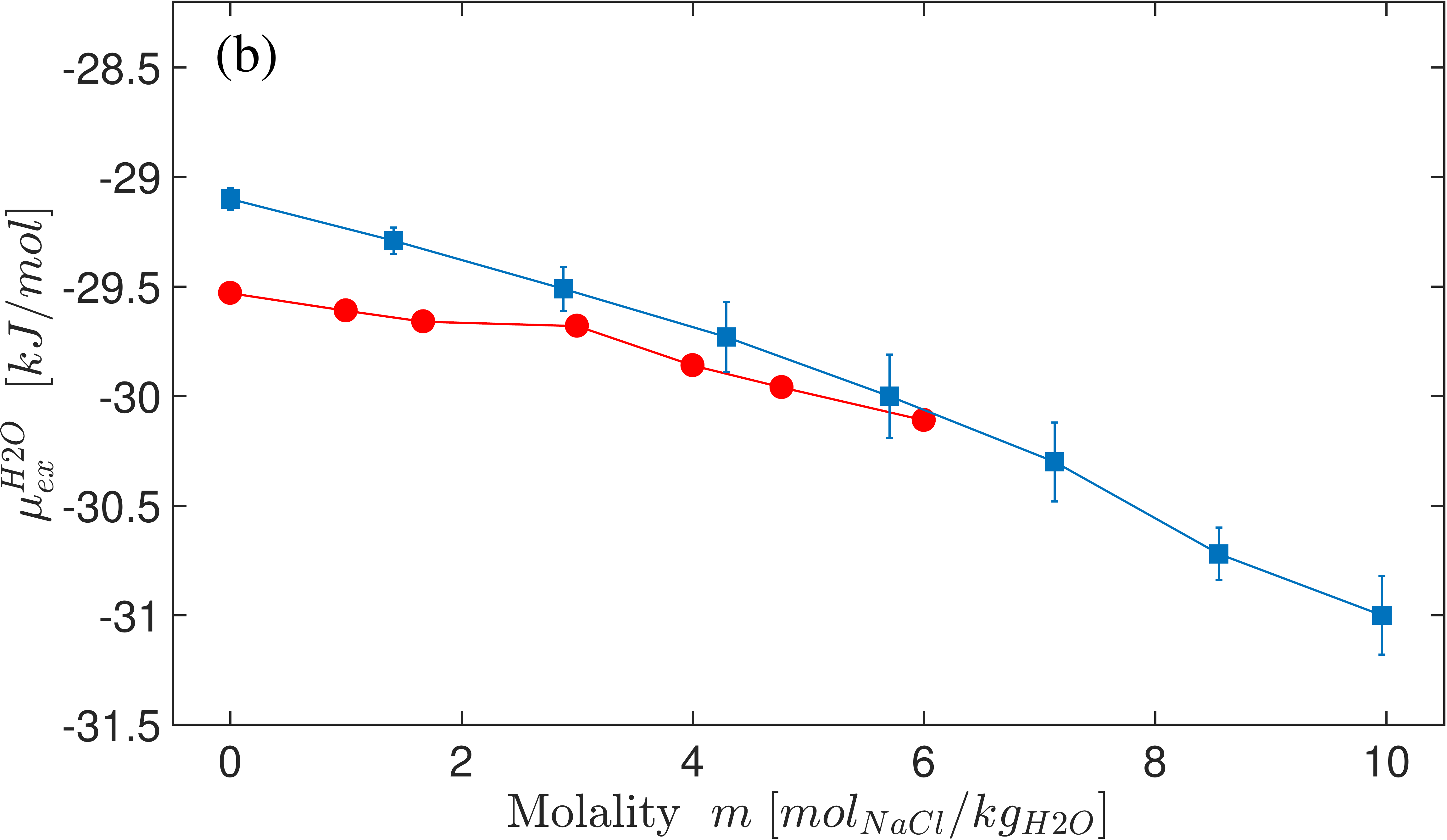}
\caption{Excess chemical potential of water molecules $\mu_{ex}^{H2O}$ (a) and molecular NaCl $\mu_{ex}^{NaCl}$ (b) as computed for different salt concentrations. The results obtained with the {\tt SPARTIAN} and BAR \cite{JCP142_044507_2015} methods are represented by blue squares and red circles, respectively.} \label{Fig3}
\end{figure}

Hence, after applying this corrections we have compared directly the results computed with our method and with those reported in Ref. \citenum{JCP142_044507_2015} using BAR \cite{BENNETT1976245,PRL91-140601-2003,JCP122-144107-2005,JCTC7-4115-2011}.
It is apparent from Fig. \ref{Fig3} that our results for $\mu_{ex}^{NaCl}$ and $\mu_{ex}^{H_{2}O}$ are both in excellent agreement with such reported values. Furthermore, we report here, for the first time to our knowledge, values of excess chemical potentials for this system for molalities larger than 7 \cite{JCP136_244508_2012}. These results show well--defined trends, indicating that upon increasing NaCl concentration the addition of a further solute molecule becomes energetically less favourable, roughly 10 kJ/mol in the range 6-10, whereas in the same range it is slightly more favourable to add another water molecule to the system (1 kJ/mol).

\section*{Conclusions}
\label{conclusions}

The chemical potential is of central importance for the comprehension of the physico-chemical properties of a substance and the capacity to manipulate it for scientific and industrial purposes. The vast majority of computational methods devised to calculate chemical potentials rely on periodic attempts to insert a test particle into the system. For dense liquids and highly concentrated liquid mixtures, this procedure is inefficient, perhaps in some cases unfeasible, and the use of enhanced sampling techniques or the design of alternative methods becomes crucial. 

In this work we presented a method, spatially resolved thermodynamic integration, or {\tt SPARTIAN}, which introduces a different perspective based on the Hamiltonian adaptive resolution framework. Here, the target system is physically separated from a reservoir of ideal gas particles by a hybrid region where molecules change resolution, from atomistic to ideal gas and vice versa, on the fly. To ensure a uniform density profile of the whole system, free energy compensations are parameterised and applied to the molecules present in the hybrid region. Under such conditions, the system reaches thermal equilibrium and the chemical potential of both target system and ideal gas reservoir equates. Therefore, the free energy compensations are identified with the difference in chemical potential between the two representations, which is precisely the excess chemical potential of the target system. 

The method is efficient and of general applicability, as demonstrated by the reported results on pure and multi-component Lennard Jones liquids, pure water, and aqueous solutions of sodium chloride. The values of the excess chemical potential computed for the various species under examination are consistent with the data in the literature where available. For those regions of concentration that remain out of the scope for most established techniques, the proposed strategy has proven especially capable of providing results in line with the trend indicated by other methods. This observation suggests that the increased molecular density represents a vantage point for the method, which avoids the necessity to perform artificial particle insertions and profits of the large number of molecules to improve the convergence of statistical averages.

The {\tt SPARTIAN} strategy reported in this work thus offers a novel, effective, and versatile instrument to compute the excess chemical potential of liquids and liquid mixtures. The method is particularly well-suited to use in cases where the density of the liquid or the concentration of solute in the mixture are high. This constitutes a significant advantage over already available techniques and paves the way 
for a broad range of applications where the accurate determination of chemical potentials is of central importance. 

\begin{acknowledgements}
We thank Claudio Perego, Omar Valsson and Debashish Mukherji for a critical reading of the manuscript. MH, RP and KK acknowledge financial support under the project SFB-TRR146 of the Deutsche Forschungsgemeinschaft. RCH gratefully acknowledges the Alexander von Humboldt Foundation for financial support. MH thanks Paolo Raiteri and Julian D. Gale for fruitful discussions on the energetic comparison between Coulomb and DSF potentials. All authors are indebted with Douglas Murphy for insightful and valuable suggestions on the composition of the manuscript.
\end{acknowledgements}

%

\end{document}